\documentclass[prd,reprint,floatfix,nofootinbib,longbibliography]{revtex4-2}

\usepackage{amsmath,amssymb}
\usepackage{capt-of}
\usepackage{graphicx}

\newcommand{\qnmTableSetup}{%
\centering
\scriptsize
\setlength{\tabcolsep}{2pt}%
\renewcommand{\arraystretch}{1.02}%
}

\newcommand{\qnmCaption}[2]{%
Scalar quasinormal frequencies $\omega$ for $\ell=#1$ and $p_2=#2$ in the parity-symmetric black-hole background with $M=1$. Columns 3 and 4 list the Pad\'e-improved 16th- and 14th-order WKB estimates, and the last column gives their relative difference in percent.}

\newcommand{\qnmHeader}{%
\hline\hline
$p_1$ & $\mu_s$ & \shortstack{WKB16--Pad\'e\\($\tilde m=8$)} & \shortstack{WKB14--Pad\'e\\($\tilde m=7$)} & \shortstack{Rel.\\diff. (\%)} \\
\hline
}

\begin{document}

\title{Massive Scalar Quasinormal Modes, Greybody Factors, and Absorption Cross Section of a Parity-Symmetric Beyond-Horndeski Black Hole}

\author{S.~V.~Bolokhov}
\email{bolokhov-sv@rudn.ru}
\affiliation{RUDN University, 6 Miklukho-Maklaya Street, Moscow, 117198, Russian Federation}

\begin{abstract}
We study quasinormal modes, greybody factors, and the absorption cross section of a massive scalar field in the asymptotically flat parity-symmetric beyond-Horndeski black-hole background. The scalar mass raises the asymptotic level of the effective potential and can eliminate its barrier peak, thereby changing both the ringing spectrum and the scattering characteristics relative to the massless case. Using Pad\'e-improved high-order WKB calculations together with time-domain evolution, we find that the damping rate decreases strongly as the field mass increases, indicating the approach to long-lived quasi-resonant states for representative parameter families. At the same time, in the large-mass regime these weakly damped modes become progressively harder to isolate in the time domain, because the oscillatory massive tails are expected to dominate on the Koyama--Tomimatsu scale $\mu_s t\gg \mu_s M$, which is comparatively early when $M=1$ and $\mu_s$ is not small. The time-domain profiles also exhibit the transition from quasinormal ringing to an oscillatory late-time tail. Interpreting the same effective potential semiclassically, we show that increasing the scalar mass suppresses low-frequency transmission and shifts the onset of efficient absorption to higher frequencies, while larger deviations from the Schwarzschild limit enhance the absorption cross section. These results show that the competition between long-lived modes and rapidly dominant massive tails makes the massive sector an especially subtle and sensitive probe of the interplay between field mass and geometric deformation in this class of black holes.
\end{abstract}

\maketitle
\setlength{\parindent}{0pt}
\setlength{\parskip}{1\baselineskip}

\section{Introduction}

Quasinormal modes provide one of the most efficient ways to encode the response of a black hole to external disturbances. Their complex frequencies govern the damped ringing seen after the initial transient stage and connect the geometry of the background spacetime with observable wave signals~\cite{Kokkotas:1999bd,Berti:2009kk,Konoplya:2011qq,Nollert:1999ji}. In modified-gravity settings this connection is especially informative, because even a relatively simple test field can reveal how changes in the metric alter the characteristic oscillation spectrum.

At the same time, the same effective potential that governs ringdown also controls the scattering of waves by the black-hole exterior. When radiation is sent from infinity, part of it is reflected by the curvature barrier and part of it penetrates toward the horizon. The corresponding transmission probabilities are the greybody factors, whereas the absorption cross section measures the effective area with which the black hole captures the incoming field. These quantities are physically important for two complementary reasons: they determine how near-horizon Hawking emission is filtered on its way to infinity, and they describe the classical absorbing properties of the geometry as seen by external probes.

A massive scalar field is an especially informative test of the geometry because its nonzero mass influences not only the peak of the effective barrier but also the far-region behavior of the wave equation. In practice, this introduces an additional physical scale and turns the spectrum into a more delicate probe of the competition between local curvature and asymptotic propagation. One therefore expects the massive sector to reveal features that remain hidden in the massless limit, including threshold effects near $\omega\sim\mu_s$, very slowly damped ringing in part of parameter space, and a qualitatively different low-frequency scattering pattern. For this reason, extending the analysis from massless perturbations to massive ones is not merely a technical generalization but a necessary step toward a fuller characterization of the background.

Massive-field spectra are also interesting for broader physical reasons. Effective mass terms can arise from extra-dimensional constructions in which bulk physics is projected onto brane perturbations~\cite{Seahra:2004fg}, while massive gravitational degrees of freedom are relevant in current discussions of ultra-low-frequency gravitational signals~\cite{Konoplya:2023fmh,NANOGrav:2023hvm}. A further motivation is the rich spectral phenomenology already uncovered in representative studies of massive perturbations~\cite{Konoplya:2004wg,Konoplya:2018qov,Konoplya:2017tvu,Zhidenko:2006rs,Konoplya:2005hr,Ohashi:2004wr,Zhang:2018jgj,Aragon:2020teq,Ponglertsakul:2020ufm,Gonzalez:2022upu,Burikham:2017gdm}. In particular, tuning the field mass may produce quasi-resonant states with exceptionally small damping, a phenomenon first identified in early analyses and later observed for different spins, black-hole geometries, and even horizonless compact objects~\cite{Ohashi:2004wr,Konoplya:2004wg,Konoplya:2005hr,Fernandes:2021qvr,Konoplya:2017tvu,Percival:2020skc,Zhidenko:2006rs,Zinhailo:2018ska,Churilova:2020bql,Bolokhov:2023bwm,Lutfuoglu:2026uzy,Lutfuoglu:2026gis,Lutfuoglu:2026xlo,Lutfuoglu:2026fpx,Lutfuoglu:2025kqp,Lutfuoglu:2025eik,Lutfuoglu:2025qkt,Lutfuoglu:2025bsf,Lutfuoglu:2025hwh,Lutfuoglu:2025hjy,Skvortsova:2026unq,Skvortsova:2025cah,Skvortsova:2024eqi,Bolokhov:2026dzn,Bolokhov:2024bke,Bolokhov:2023ruj,Dubinsky:2025wns,Dubinsky:2025bvf,Malik:2025czt,Churilova:2019qph}. At the same time, the late-time signal is modified in an equally characteristic way, since the standard power-law tail can be replaced by oscillatory decay~\cite{Jing:2004zb,Koyama:2001qw,Moderski:2001tk,Rogatko:2007zz,Koyama:2001ee,Koyama:2000hj,Gibbons:2008gg,Gibbons:2008rs,Dubinsky:2024jqi}, and effective mass terms may even emerge for otherwise massless fields in magnetized black-hole backgrounds~\cite{Konoplya:2007yy,Konoplya:2008hj,Wu:2015fwa}. At the same time, quasi-resonant behavior is not universal, so determining when such long-lived modes do or do not occur is itself part of the problem~\cite{Zinhailo:2024jzt,Konoplya:2005hr}.

The black-hole geometry considered here comes from the parity-symmetric branch of beyond-Horndeski gravity derived in Ref.~\cite{Bakopoulos:2022nqd} and used recently in studies of greybody factors and massless quasinormal modes~\cite{Antoniou:2025fpc}. More specifically, the massless-field analysis of the same background may be regarded as the immediate precursor of the present work, since here we keep the geometry and observables fixed but extend the discussion from massless probes to a massive scalar field. This continuation is physically important because the additional scale $\mu_s$ changes both the quasinormal spectrum and the scattering characteristics that were already identified in the massless case. The purpose of the present draft is therefore to set up both the resonant and scattering aspects of the massive scalar problem in a unified form. To that end we collect the essential ingredients of the background, derive the radial equation and effective potential for a field of mass $\mu_s$, explain how the same potential yields quasinormal modes, greybody factors, and the absorption cross section, and summarize the numerical approaches most natural in this setting, namely higher-order WKB methods, barrier-scattering techniques, and characteristic time-domain evolution.

The remainder of the paper is organized as follows. In Sec.~II we briefly review the parity-symmetric beyond-Horndeski background and fix the notation for the metric and deformation parameters. Section~III introduces the massive scalar wave equation and the corresponding effective potential. In Sec.~IV we summarize the Pad\'e-improved higher-order WKB approach, while Sec.~V discusses the characteristic time-domain evolution and the extraction of dominant frequencies. The main quasinormal-mode results are presented in Sec.~VI, including the behavior of the damping rate, the onset of quasi-resonant regimes, and the role of late-time massive tails. Section~VII is devoted to the greybody factors and the absorption cross section, interpreted in the present QNM-correspondence framework. Finally, Sec.~VIII summarizes the main conclusions and outlines possible extensions of the analysis.

\section{Essentials of the theory and metric}

We work in the shift-symmetric, parity-symmetric sector of beyond-Horndeski theory in four spacetime dimensions~\cite{Bakopoulos:2022nqd}. The gravitational action can be written as
\begin{equation}
S = \int d^4x\,\sqrt{-g}\,\bigl(\mathcal{L}_2 + \mathcal{L}_4 + \mathcal{L}^{\rm bH}_4\bigr),
\end{equation}
where
\begin{align}
\mathcal{L}_2 &= G_2(X), \\
\mathcal{L}_4 &= G_4(X)R + G_{4X}(X)\Bigl[(\Box\phi)^2 - \phi_{\mu\nu}\phi^{\mu\nu}\Bigr], \\
\mathcal{L}^{\rm bH}_4 &= F_4(X)\,\varepsilon^{\mu\nu\rho}{}_{\sigma}\,\varepsilon^{\mu'\nu'\rho'\sigma}\,\phi_{\mu}\phi_{\mu'}\phi_{\nu\nu'}\phi_{\rho\rho'}.
\end{align}
Here $X\equiv -\nabla_{\mu}\phi\nabla^{\mu}\phi/2$, while the odd-parity functions $G_3$, $G_5$, and $F_5$ are absent in the sector of interest.

For the asymptotically flat black-hole branch relevant to quasinormal-mode calculations, the spacetime metric may be written in Schwarzschild-like coordinates as~\cite{Bakopoulos:2022nqd,Antoniou:2025fpc}
\begin{equation}
\label{eq:ms-metric}
ds^2 = -A(r)dt^2 + \frac{dr^2}{A(r)} + r^2(d\theta^2 + \sin^2\theta\,d\varphi^2),
\end{equation}
with metric function
\begin{equation}
\label{eq:ms-A}
A(r) = 1 + p_2\frac{\arctan(p_1 r)}{p_1 r} - \frac{2M}{r}.
\end{equation}
The deformation parameters are related to the underlying couplings by
\begin{equation}
p_1 = \sqrt{\frac{\epsilon}{\delta}} > 0,
\qquad
p_2 = \frac{(\beta-\delta\zeta)^2}{8\delta\mu},
\end{equation}
where $M$ is the black-hole mass parameter and the symbols on the right-hand side refer to the constants of the exact beyond-Horndeski solution~\cite{Bakopoulos:2022nqd}. The event horizon is located at the largest positive root of
\begin{equation}
A(r_h)=0.
\end{equation}

Several basic properties of the geometry will be used repeatedly below. First, the spacetime is asymptotically flat because $A(r)\to 1$ as $r\to \infty$. Second, the Schwarzschild limit is recovered either by sending $p_2\to 0$ or by taking $p_1\to \infty$ at fixed $p_2$ and finite radius, since in both cases the arctangent deformation disappears. Third, for large $r$ one has
\begin{equation}
A(r) = 1 - \frac{2M - \pi p_2/(2p_1)}{r} - \frac{p_2}{p_1^2 r^2} + \mathcal{O}(r^{-4}),
\end{equation}
so the far zone resembles a Reissner--Nordstr\"om-type correction whenever $p_2\neq 0$. In the perturbative analysis of this paper the metric~\eqref{eq:ms-metric} is treated as a fixed background.

\section{Wave equation and effective potential}

Let $\Phi$ be a minimally coupled scalar field of mass $\mu_s$. Its dynamics is governed by the massive Klein--Gordon equation,
\begin{equation}
\label{eq:ms-KG}
\left(\nabla_{\mu}\nabla^{\mu} - \mu_s^2\right)\Phi = 0,
\end{equation}
where $\mu_s$ should not be confused with the coupling parameter $\mu$ appearing in the underlying gravitational theory. With the usual spherical decomposition \cite{Carter1968HJ,Carter1968Kerr,Konoplya:2018arm},
\begin{equation}
\label{eq:ms-separation}
\Phi(t,r,\theta,\varphi)
= e^{-i\omega t}\,Y_{\ell m}(\theta,\varphi)\,\frac{\Psi(r)}{r},
\qquad \ell = 0,1,2,\dots,
\end{equation}
and with the tortoise coordinate defined by
\begin{equation}
\label{eq:ms-tortoise}
\frac{dr_*}{dr} = \frac{1}{A(r)},
\end{equation}
one obtains the radial master equation
\begin{equation}
\label{eq:ms-master}
\frac{d^2\Psi}{dr_*^2} + \bigl[\omega^2 - V_{\ell}(r)\bigr]\Psi = 0.
\end{equation}
The effective potential takes the compact form
\begin{equation}
\label{eq:ms-potential-compact}
V_{\ell}(r) = A(r)\left[\frac{\ell(\ell+1)}{r^2} + \frac{A'(r)}{r} + \mu_s^2\right],
\end{equation}
with
\begin{equation}
\label{eq:ms-Aprime}
A'(r) = p_2\left[\frac{1}{r\bigl(1+p_1^2r^2\bigr)} - \frac{\arctan(p_1r)}{p_1r^2}\right] + \frac{2M}{r^2}.
\end{equation}
After substituting Eq.~\eqref{eq:ms-Aprime} into Eq.~\eqref{eq:ms-potential-compact}, the potential can also be written as
\begin{equation}
\label{eq:ms-potential}
\begin{aligned}
V_{\ell}(r) = A(r)\Biggl[&\mu_s^2 + \frac{\ell(\ell+1)}{r^2} + \frac{2M}{r^3} \\
&+ \frac{p_2}{r^2(1+p_1^2r^2)} - \frac{p_2}{p_1r^3}\arctan(p_1r)\Biggr].
\end{aligned}
\end{equation}

The two limits of the potential are qualitatively important. Near the event horizon one has $A(r_h)=0$, so $V_{\ell}(r)\to 0$ as $r_*\to -\infty$. At spatial infinity the geometry approaches flat space and the potential tends to the nonzero constant
\begin{equation}
V_{\ell}(r) \longrightarrow \mu_s^2, \qquad r\to\infty.
\end{equation}
This asymptotic plateau is the main difference from the massless case. In particular, the far-zone radial equation becomes
\begin{equation}
\frac{d^2\Psi}{dr_*^2} + \chi^2\Psi \simeq 0,
\qquad
\chi \equiv \sqrt{\omega^2-\mu_s^2}.
\end{equation}

Equation~\eqref{eq:ms-potential} shows explicitly how the field mass and the background deformation compete. The mass term lifts the entire outer part of the potential, whereas the geometric parameters $p_1$ and $p_2$ reshape the barrier itself through the arctangent contribution. In the limit $\mu_s\to 0$ one recovers the massless scalar potential already discussed for this background~\cite{Antoniou:2025fpc}.

\begin{figure*}[t]
    \centering
    \includegraphics[width=\textwidth]{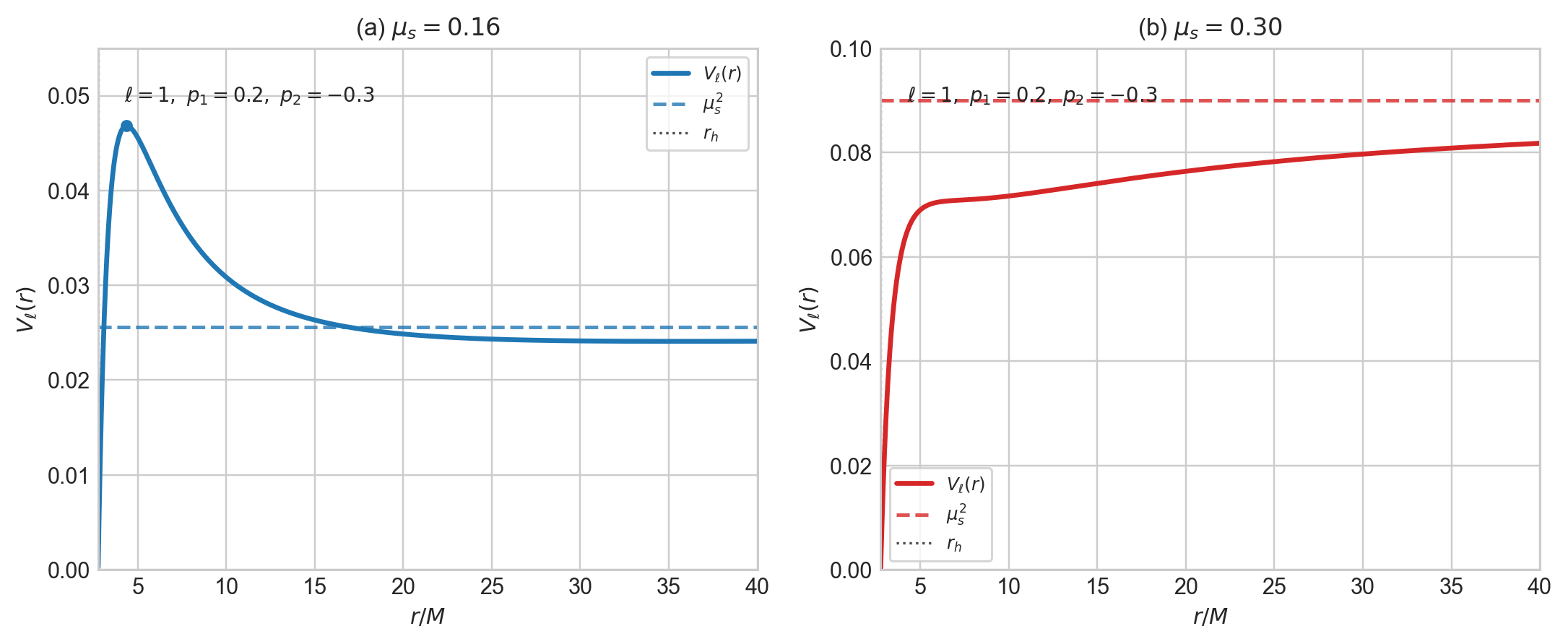}
    \caption{Representative effective potentials $V_{\ell}(r)$ for a massive scalar field with $\ell=1$, $M=1$, $p_1=0.2$, and $p_2=-0.3$. Panel (a) corresponds to $\mu_s=0.16$, for which the potential retains a standard single-barrier profile and approaches the asymptotic plateau $V_{\ell}(r)\to\mu_s^2$ at large $r$. Panel (b) corresponds to $\mu_s=0.30$, for which the barrier has disappeared and the potential increases monotonically toward the corresponding plateau. The dashed horizontal lines indicate $\mu_s^2$, while the dotted vertical lines mark the event horizon.}
    \label{fig:ms-potential-profiles}
\end{figure*}

These qualitative changes are illustrated in Fig.~\ref{fig:ms-potential-profiles} for the representative choice $\ell=1$, $M=1$, $p_1=0.2$, and $p_2=-0.3$. For $\mu_s=0.16$ the effective potential still exhibits a single barrier peak above the asymptotic plateau $\mu_s^2$, whereas for $\mu_s=0.30$ the local maximum has disappeared and the potential rises monotonically from the horizon toward its limiting value. For this particular parameter set, the crossover between the two behaviors occurs near $\mu_s\approx0.239$, which marks the point where the usual single-barrier WKB picture ceases to be the natural description.

\section{WKB method}

For quasinormal modes one imposes a purely ingoing wave at the horizon and a purely outgoing wave at infinity,
\begin{equation}
\Psi \sim e^{-i\omega r_*}, \qquad r_*\to -\infty,
\end{equation}
\begin{equation}
\Psi \sim e^{+i\chi r_*}, \qquad r_*\to +\infty,
\end{equation}
with the branch of $\chi$ chosen so that outgoing propagation is selected. If $\omega^2$ lies below the mass threshold, $\chi$ becomes imaginary and the same master equation naturally crosses over to the quasibound-state problem.

For modes governed by a single dominant potential barrier, the radial equation~\eqref{eq:ms-master} is well suited to the higher-order WKB approach developed by Iyer and Will and subsequently refined in several directions~\cite{Iyer:1986np,Konoplya:2003ii,Matyjasek:2017psv,Matyjasek:2019eeu,Konoplya:2026wkb}. Let $r_0$ denote the point where the potential reaches its maximum and let
\begin{equation}
V_0 = V_{\ell}(r_0),
\qquad
V_0'' = \left.\frac{d^2V_{\ell}}{dr_*^2}\right|_{r_0}.
\end{equation}
At WKB order $N$ the quasinormal frequencies are determined from the quantization condition
\begin{equation}
\label{eq:ms-wkb}
\frac{i\bigl(\omega^2-V_0\bigr)}{\sqrt{-2V_0''}} - \sum_{k=2}^{N}\Lambda_k
= n+\frac{1}{2},
\end{equation}
where $n$ is the overtone number and the correction terms $\Lambda_k$ depend on higher derivatives of the potential at the peak. In practice it is convenient to evaluate these derivatives numerically after expanding the metric and potential locally around $r_0$, which avoids cumbersome analytic differentiation of the arctangent profile.

A closely related semiclassical treatment can also be used on the scattering side of the problem. Instead of imposing the resonance condition, one evaluates the reflection and transmission coefficients of the same barrier potential, which then determine the partial greybody factors and, after summing over $\ell$, the absorption cross section~\cite{Konoplya:2019hlu,Konoplya:2026wkb}. Thus the WKB framework is useful here not only for quasinormal frequencies but also for the frequency-dependent transmission properties of the massive scalar field.

For the present geometry, high-order WKB calculations are best used together with Pad\'e resummation of the truncated series for $\omega^2$~\cite{Matyjasek:2019eeu,Konoplya:2026wkb,Hatsuda:2019borel}. This step is especially helpful when the scalar mass $\mu_s$ is large enough to broaden the barrier or to create a shallow outer plateau, because the raw asymptotic series may then converge slowly. A practical strategy is to monitor a diagonal or near-diagonal sequence of Pad\'e approximants, compare neighboring truncation orders, and regard stable clusters of values as the most reliable estimates. The WKB at various orders was applied in a number of works \cite{Konoplya:2005sy,Malik:2024iky,Guo:2020caw,Konoplya:2006ar,Lutfuoglu:2026gis,Bolokhov:2025egl,Skvortsova:2024msa,Kokkotas:2010zd,Eniceicu:2019npi,Lutfuoglu:2025ohb,Konoplya:2010vz,Abdalla:2005hu,Wongjun:2019ydo,Malik:2025erb,Lutfuoglu:2025ljm,Konoplya:2009hv,Stuchlik:2025mjj,Fernando:2016ftj,Konoplya:2019ppy,Breton:2017hwe,Ishihara:2008re,Pathrikar:2025gzu,Konoplya:2007zx,Karmakar:2023cwg,Bolokhov:2025aqy,Konoplya:2023ppx,Momennia:2018hsm,Bolokhov:2025lnt,Lutfuoglu:2026gey,Malik:2024sxv}.

The massive field introduces an additional caveat that is absent in the massless problem. Since $V_{\ell}(r)$ approaches $\mu_s^2$ at infinity rather than zero, one should verify that the mode under consideration still behaves as a barrier-scattering state and not as a quasibound state near the mass threshold. Therefore, any WKB calculation should be accompanied by checks on the sign and magnitude of $\chi^2=\omega^2-\mu_s^2$ as well as by an independent time-domain test whenever the damping becomes very weak.

\section{Time-domain integration}

A complementary route to the spectrum is obtained by evolving the wave equation directly in the time domain~\cite{Gundlach:1994jp,Konoplya:2011qq,Dubinsky:2024gwo}. Starting from
\begin{equation}
\label{eq:ms-time-wave}
\left(-\frac{\partial^2}{\partial t^2}+\frac{\partial^2}{\partial r_*^2}-V_{\ell}(r)\right)\Psi(t,r_*)=0,
\end{equation}
one introduces the null coordinates
\begin{equation}
u = t-r_*, \qquad v = t+r_*.
\end{equation}
The master equation becomes
\begin{equation}
\label{eq:ms-null}
4\frac{\partial^2\Psi}{\partial u\,\partial v} + V_{\ell}(r)\Psi = 0.
\end{equation}
A standard characteristic discretization on a null grid with points $S=(u,v)$, $E=(u,v+\Delta)$, $W=(u+\Delta,v)$, and $N=(u+\Delta,v+\Delta)$ gives
\begin{equation}
\label{eq:ms-gpp}
\Psi_N = \Psi_W + \Psi_E - \Psi_S - \frac{\Delta^2}{8}V_{\ell}(S)\bigl(\Psi_W+\Psi_E\bigr) + \mathcal{O}(\Delta^4).
\end{equation}
One typically starts from a Gaussian pulse on one initial null segment together with trivial data on the other,
\begin{equation}
\Psi(u,u_0)=\exp\!\left[-\frac{(u-u_c)^2}{2\sigma^2}\right],
\qquad
\Psi(u_0,v)=0,
\end{equation}
and then records the signal at a fixed observation radius.

The dominant quasinormal frequencies can be extracted from the ringing stage by fitting the waveform to a sum of damped exponentials,
\begin{equation}
\label{eq:ms-prony}
\Psi(t,r_*^{\rm obs}) \approx \sum_{j=1}^{p} C_j e^{-i\omega_j (t-t_0)}.
\end{equation}
For a massive field this procedure requires some care, because the late-time signal may contain oscillatory tails influenced by the mass term. The most reliable fit window is therefore an intermediate interval: late enough for the prompt response to have died away, but not so late that the tail dominates over the exponentially damped ringing. Once this window is identified, the extracted frequencies provide a direct consistency check for the WKB predictions \cite{Skvortsova:2024atk,Konoplya:2024hfg,Abdalla:2012si,Bolokhov:2023bwm,Malik:2023bxc,Churilova:2021tgn,Bolokhov:2024ixe,Arbelaez:2025gwj,Varghese:2011ku,Dubinsky:2024nzo,Skvortsova:2023zmj,Bolokhov:2026eqf,Lutfuoglu:2025hjy,Konoplya:2013sba,Skvortsova:2024wly,Malik:2024nhy,Arbelaez:2026eaz,Bolokhov:2023dxq,Dubinsky:2025fwv}.

\begin{figure}[t]
    \centering
    \includegraphics[width=\linewidth]{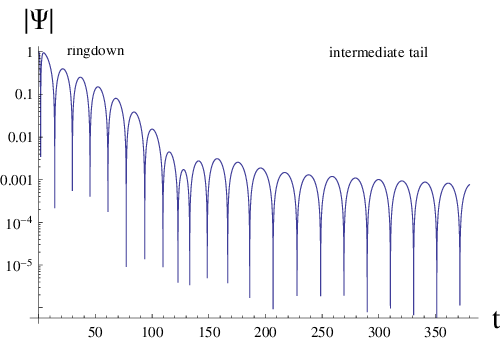}
    \caption{Semilogarithmic time-domain profile of the absolute value of the massive scalar perturbation, $|\Psi|$, for $\ell=1$, $M=1$, $p_1=0.2$, $p_2=-0.3$, and $\mu_s=0.16$, evaluated at a fixed observer position. The waveform shows a clear quasinormal-ringing regime up to approximately $t\approx120$, after which an intermediate oscillatory tail becomes visible. A Prony fit to the ringing stage gives $\omega = 0.204411 - 0.0456868 i$, while the corresponding WKB16--Pad\'e estimate is $\omega = 0.204362 - 0.045567 i$. The differences in the real and imaginary parts are about $0.024\%$ and $0.263\%$, respectively, so the overall discrepancy is of order $10^{-4}$ in absolute units.}
    \label{fig:ms-td-profile}
\end{figure}

Figure~\ref{fig:ms-td-profile} illustrates the expected transition between the exponentially damped quasinormal ringing and the later-time massive-field tail. The close agreement between the Prony and WKB frequencies confirms that the interval before $t\approx120$ is an appropriate fitting window for the dominant mode.

\section{Quasinormal modes}

The damping rate $\gamma=-\operatorname{Im}(\omega)$ decreases markedly as the scalar-field mass increases for several combinations of $\ell$ and $p_1$. This tendency is illustrated in Fig.~\ref{fig:ms-damping-threshold} for representative data sets with fixed $p_2=-0.6$, where the last available points are extrapolated to estimate the critical value of $\mu_s$ at which the damping rate tends to zero.

\begin{center}
\includegraphics[width=\columnwidth]{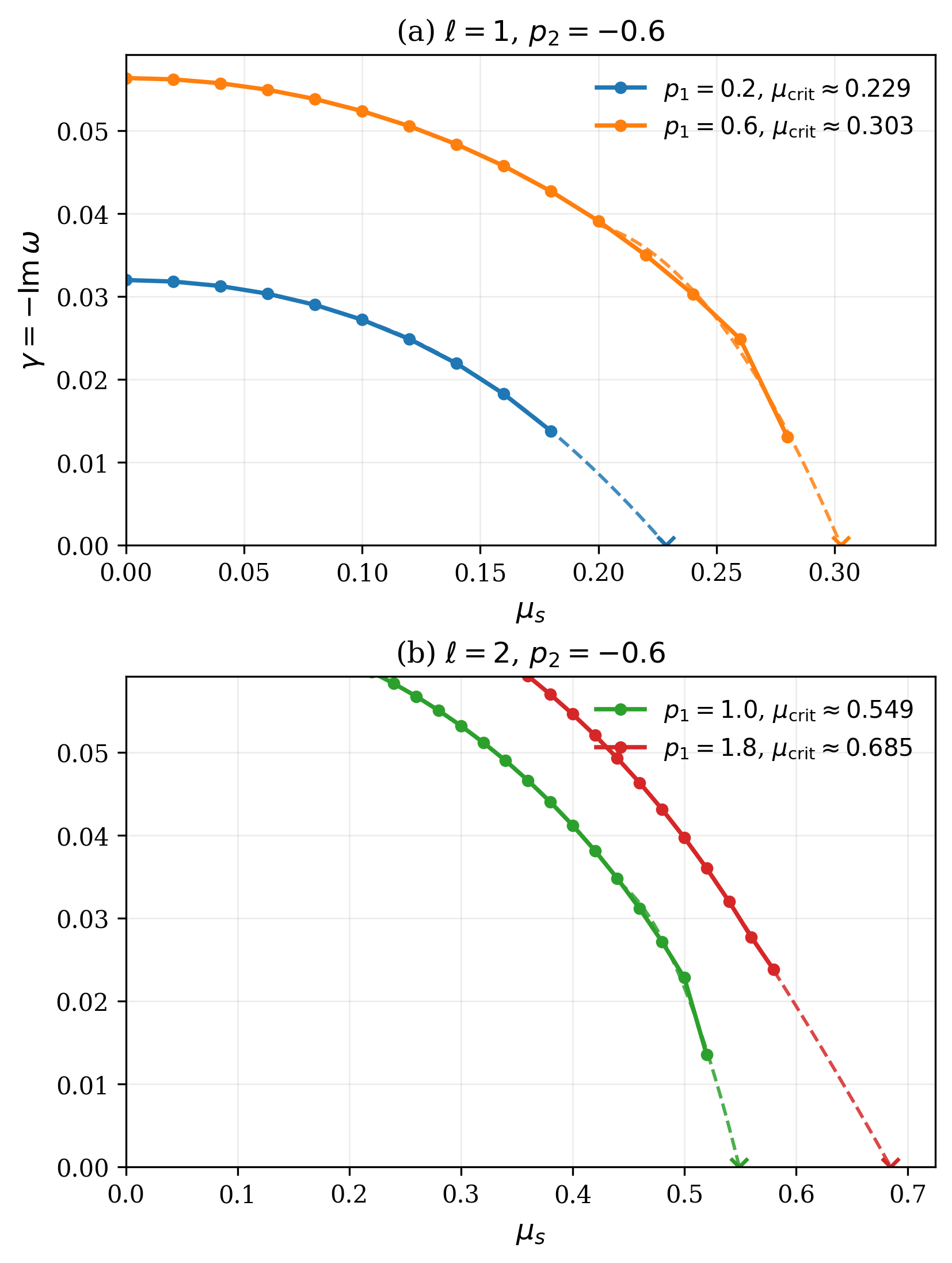}
\captionof{figure}{Damping rate $\gamma=-\operatorname{Im}(\omega)$ as a function of the scalar-field mass $\mu_s$ for representative quasinormal modes with $p_2=-0.6$. Panel (a) shows the $\ell=1$ data for $p_1=0.2$ and $p_1=0.6$, while panel (b) shows the $\ell=2$ data for $p_1=1.0$ and $p_1=1.8$. The markers are taken from the WKB16 results listed in the tables, and the dashed curves are quadratic extrapolations of the last five available points in each series. The corresponding zero-damping estimates are $\mu_{\rm crit}\approx 0.229$, $0.303$, $0.549$, and $0.685$, respectively.}
\label{fig:ms-damping-threshold}
\end{center}

Before drawing physical conclusions from the spectra, it is important to emphasize what the comparison between the WKB16--Pad\'e and WKB14--Pad\'e columns can and cannot tell us. This inter-order spread is not a rigorous error bar, but it is a very useful internal accuracy indicator: when two high orders agree to many displayed digits, the corresponding mode is typically robust, while a visible separation between the two orders warns that the mode lies in a more delicate regime. Read in this way, Tables~\ref{tab:ms-qnm-l0-p2m03}, \ref{tab:ms-qnm-l1-p2m03}, \ref{tab:ms-qnm-l0-p2m06}, \ref{tab:ms-qnm-l1-p2m06}, and \ref{tab:ms-qnm-l2-p2m06} show a clear hierarchy of reliability.

The most difficult cases are the $\ell=0$ modes collected in Tables~\ref{tab:ms-qnm-l0-p2m03} and \ref{tab:ms-qnm-l0-p2m06}. There the median relative differences between WKB14 and WKB16 are already at the level of about $0.38\%$ and $0.50\%$, respectively, and several isolated entries reach the few-percent level, with maxima of about $6.85\%$ in Table~\ref{tab:ms-qnm-l0-p2m03} and $11.7\%$ in Table~\ref{tab:ms-qnm-l0-p2m06}. This is qualitatively what one expects: the lowest multipole is always the least favorable regime for WKB, and the agreement deteriorates further when the scalar mass reshapes the barrier into a shallow structure. By contrast, the $\ell=1$ and $\ell=2$ tables are much more stable. In Table~\ref{tab:ms-qnm-l1-p2m03} every displayed entry remains below $1\%$, with a mean discrepancy of about $5.8\times 10^{-2}\%$; in Table~\ref{tab:ms-qnm-l1-p2m06} the mean discrepancy is about $6.2\times 10^{-2}\%$; and in Table~\ref{tab:ms-qnm-l2-p2m06} the agreement is even better, with a mean of about $2.2\times 10^{-2}\%$ and a maximum of only $0.332\%$. Therefore, over most of the displayed parameter range the $\ell\ge1$ frequencies appear quantitatively reliable to much better than one percent, whereas the $\ell=0$ values should be read more cautiously whenever the two WKB orders begin to split.

Figure~\ref{fig:ms-td-profile} provides an independent time-domain check of this internal accuracy estimate. For the representative mode with $\ell=1$, $p_1=0.2$, $p_2=-0.3$, and $\mu_s=0.16$, the Prony fit reproduces the WKB16--Pad\'e result with differences of only $0.024\%$ in ${\rm Re}(\omega)$ and $0.263\%$ in $|{\rm Im}(\omega)|$. This agreement is fully consistent with the tiny inter-order differences seen for the neighboring entries in Table~\ref{tab:ms-qnm-l1-p2m03}, and it supports the view that, as long as a clean ringdown window is present, the WKB values do capture the dominant mode with high accuracy.

\begin{figure}[t]
    \centering
    \includegraphics[width=0.92\linewidth]{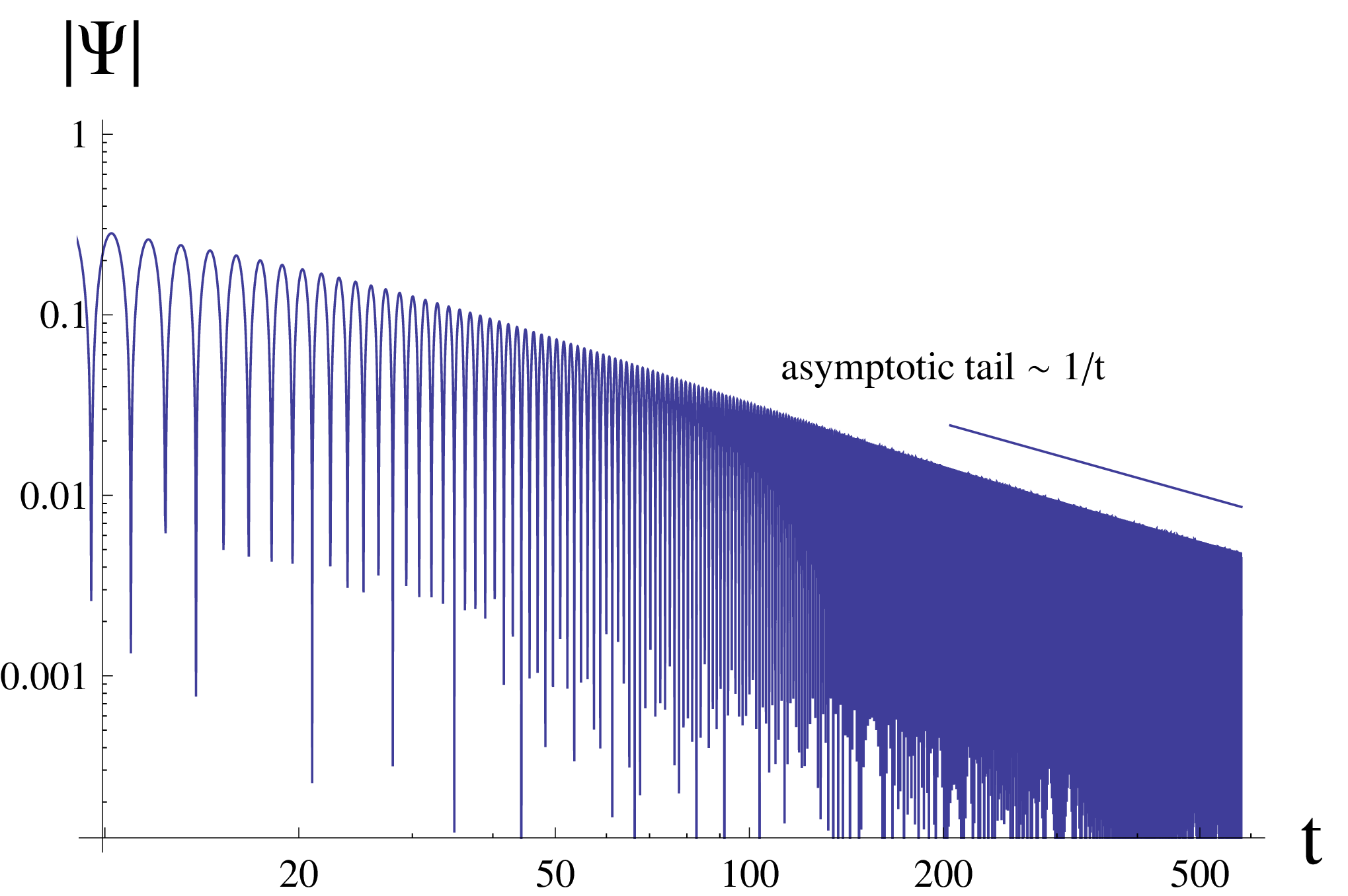}
    \caption{Oscillatory late-time tail of the massive scalar perturbation for $\ell=1$, $M=1$, $p_2=-0.3$, $p_1=0.2$, and $\mu_s=0.3$. After the quasinormal ringing stage, the signal develops a clear oscillatory tail whose upper envelope is well approximated in the displayed asymptotic regime by the power law $|\Psi|\propto t^{-1}$.}
    \label{fig:ms-tail-power}
\end{figure}

The newly added tail profile in Fig.~\ref{fig:ms-tail-power} shows that, for this representative choice of parameters, the late-time evolution crosses over from the exponentially damped quasinormal signal to an oscillatory massive tail whose envelope decays approximately as $t^{-1}$. In that respect the asymptotic behavior is qualitatively similar to the familiar oscillatory late-time tails discussed for massive perturbations on Schwarzschild and Reissner--Nordstr\"om backgrounds in Refs.~\cite{Koyama:2001ee,Koyama:2001qw}. This observation also reinforces the practical point made above: the Prony fit must be restricted to the earlier ringdown interval, because once the oscillatory power-law tail takes over the waveform is no longer described by a small superposition of damped quasinormal exponentials. This limitation becomes especially important precisely in the regime where one expects quasi-resonant behavior. The classic analysis of Koyama and Tomimatsu implies that, for small $\mu_s M$, the asymptotic massive tail takes over only at very late times, $\mu_s t\gg 1/(\mu_s M)^2$, but for larger $\mu_s M$ the onset scale is instead $\mu_s t\gg \mu_s M$~\cite{Koyama:2001ee,Koyama:2001qw}. Therefore, when the field mass is large enough to make the QNMs very long lived, the tail can dominate after only a comparatively short evolution, so the weakly damped mode may exist in the spectrum but still fail to appear as a clean, long-lasting ringing stage in the time-domain profile.

The physically most interesting trend in the entire data set is the strong suppression of the damping rate as the scalar mass increases. This is visible directly in Table~\ref{tab:ms-qnm-l0-p2m03}, Table~\ref{tab:ms-qnm-l1-p2m03}, Table~\ref{tab:ms-qnm-l0-p2m06}, Table~\ref{tab:ms-qnm-l1-p2m06}, and Table~\ref{tab:ms-qnm-l2-p2m06}, and it is summarized most clearly in Fig.~\ref{fig:ms-damping-threshold}. The extrapolated zero-damping points in that figure, $\mu_{\rm crit}\approx0.229$, $0.303$, $0.549$, and $0.685$, show that for several representative sequences the imaginary part is driven toward zero in a systematic way rather than by an isolated numerical accident. The natural interpretation is that the system is approaching quasi-resonant configurations with extremely long lifetimes. At the same time, this conclusion must be stated carefully: the present tables and Fig.~\ref{fig:ms-damping-threshold} strongly indicate, but certainly do not prove, the existence of arbitrarily long-lived quasi-resonances. Near the weakest-damping points the extrapolation becomes more delicate, and exactly there the WKB method is under the greatest stress, especially for the $\ell=0$ sector.

Figure~\ref{fig:ms-potential-profiles} helps explain why the damping drops so strongly. Panel~(a) shows the familiar single-barrier situation for $\mu_s=0.16$, while panel~(b) shows that for larger $\mu_s$ the effective potential can lose its local maximum altogether and rise monotonically toward the asymptotic plateau. This deformation of the potential is precisely the mechanism that weakens the leakage of the mode and makes the damping small. It also clarifies why one should be more cautious in the near-threshold regime: once the problem is no longer controlled by a clean single barrier, the usual WKB quantization picture becomes less rigid.

A direct numerical check of the barrier maxima obtained from Eq.~\eqref{eq:ms-potential} also confirms the expected correlation between ${\rm Re}(\omega)$ and the height of the effective potential. For the $\ell=1$ and $\ell=2$ data in Tables~\ref{tab:ms-qnm-l1-p2m03}, \ref{tab:ms-qnm-l1-p2m06}, and \ref{tab:ms-qnm-l2-p2m06}, the correlation between ${\rm Re}(\omega)$ and $\sqrt{V_0}$ is extremely strong and is nearly monotonic along each fixed-$p_1$ sequence; the same tendency is still present overall in Tables~\ref{tab:ms-qnm-l0-p2m03} and \ref{tab:ms-qnm-l0-p2m06}, although with more visible low-$\ell$ irregularities near the threshold region. In other words, the qualitative rule of thumb suggested by the data is correct: the taller the effective barrier, the larger the oscillation frequency. This is not an exact identity, especially once the mass term substantially reshapes the outer part of the potential, but the trend is clearly borne out by all five tables and is fully consistent with the representative profiles displayed in Fig.~\ref{fig:ms-potential-profiles}.

\begin{table}[t]
\qnmTableSetup
\begin{tabular}{@{}c c c c c@{}}
\qnmHeader
$0.2$ & $0$ & $0.068891-0.056598 i$ & $0.069020-0.056663 i$ & $0.162\%$\\
$0.2$ & $0.04$ & $0.069515-0.054597 i$ & $0.069626-0.054693 i$ & $0.167\%$\\
$0.2$ & $0.08$ & $0.073148-0.049631 i$ & $0.073096-0.049610 i$ & $0.0631\%$\\
$0.2$ & $0.12$ & $0.084827-0.036349 i$ & $0.083050-0.031256 i$ & $5.84\%$\\
$0.2$ & $0.14$ & $0.092820-0.031598 i$ & $0.095060-0.028056 i$ & $4.27\%$\\
$0.6$ & $0$ & $0.085606-0.077055 i$ & $0.085697-0.076884 i$ & $0.169\%$\\
$0.6$ & $0.04$ & $0.086075-0.075424 i$ & $0.086148-0.075222 i$ & $0.188\%$\\
$0.6$ & $0.08$ & $0.087839-0.070825 i$ & $0.088111-0.069983 i$ & $0.785\%$\\
$0.6$ & $0.12$ & $0.089425-0.063503 i$ & $0.090195-0.062719 i$ & $1.00\%$\\
$0.6$ & $0.16$ & $0.086261-0.055575 i$ & $0.085604-0.057339 i$ & $1.83\%$\\
$0.6$ & $0.18$ & $0.121134-0.049689 i$ & $0.124041-0.044017 i$ & $4.87\%$\\
$1.$ & $0$ & $0.092647-0.085695 i$ & $0.092487-0.085487 i$ & $0.208\%$\\
$1.$ & $0.04$ & $0.093152-0.084117 i$ & $0.092902-0.083916 i$ & $0.255\%$\\
$1.$ & $0.08$ & $0.094318-0.079119 i$ & $0.097191-0.076150 i$ & $3.36\%$\\
$1.$ & $0.12$ & $0.096762-0.072754 i$ & $0.096569-0.072315 i$ & $0.397\%$\\
$1.$ & $0.16$ & $0.099180-0.061314 i$ & $0.098774-0.061419 i$ & $0.360\%$\\
$1.$ & $0.2$ & $0.140346-0.039618 i$ & $0.142788-0.033906 i$ & $4.26\%$\\
$1.4$ & $0$ & $0.096462-0.090278 i$ & $0.096368-0.089992 i$ & $0.228\%$\\
$1.4$ & $0.04$ & $0.096899-0.088807 i$ & $0.096768-0.088462 i$ & $0.281\%$\\
$1.4$ & $0.08$ & $0.098266-0.084435 i$ & $0.097070-0.081986 i$ & $2.10\%$\\
$1.4$ & $0.12$ & $0.100290-0.077487 i$ & $0.100358-0.077524 i$ & $0.0612\%$\\
$1.4$ & $0.16$ & $0.099305-0.073164 i$ & $0.099305-0.073165 i$ & $0.0006\%$\\
$1.4$ & $0.2$ & $0.103249-0.077632 i$ & $0.112038-0.078614 i$ & $6.85\%$\\
$1.8$ & $0$ & $0.098943-0.093093 i$ & $0.098862-0.092791 i$ & $0.230\%$\\
$1.8$ & $0.04$ & $0.099360-0.091656 i$ & $0.099251-0.091294 i$ & $0.280\%$\\
$1.8$ & $0.08$ & $0.100691-0.087481 i$ & $0.099552-0.085761 i$ & $1.55\%$\\
$1.8$ & $0.12$ & $0.102752-0.080353 i$ & $0.102798-0.080408 i$ & $0.0549\%$\\
$1.8$ & $0.16$ & $0.102598-0.072892 i$ & $0.102332-0.074590 i$ & $1.37\%$\\
$1.8$ & $0.2$ & $0.096240-0.066889 i$ & $0.097303-0.073171 i$ & $5.44\%$\\
$1.8$ & $0.22$ & $0.173180-0.008846 i$ & $0.175388-0.006895 i$ & $1.70\%$\\
\hline\hline
\end{tabular}
\caption{\qnmCaption{0}{-0.3}}
\label{tab:ms-qnm-l0-p2m03}
\end{table}

\begin{table}[t]
\qnmTableSetup
\begin{tabular}{@{}c c c c c@{}}
\qnmHeader
$0.2$ & $0$ & $0.188926-0.054365 i$ & $0.188926-0.054365 i$ & $0.\times 10^{\text{-4}}\%$\\
$0.2$ & $0.08$ & $0.192766-0.052254 i$ & $0.192766-0.052254 i$ & $0\%$\\
$0.2$ & $0.12$ & $0.197587-0.049535 i$ & $0.197586-0.049535 i$ & $0.0005\%$\\
$0.2$ & $0.16$ & $0.204362-0.045567 i$ & $0.204362-0.045566 i$ & $0.0004\%$\\
$0.2$ & $0.2$ & $0.213063-0.040167 i$ & $0.213070-0.040168 i$ & $0.0030\%$\\
$0.2$ & $0.24$ & $0.223578-0.033056 i$ & $0.223572-0.033193 i$ & $0.0606\%$\\
$0.2$ & $0.28$ & $0.236672-0.025450 i$ & $0.236763-0.025496 i$ & $0.0428\%$\\
$0.6$ & $0$ & $0.227207-0.072141 i$ & $0.227207-0.072141 i$ & $0\%$\\
$0.6$ & $0.08$ & $0.230780-0.070068 i$ & $0.230780-0.070068 i$ & $0\%$\\
$0.6$ & $0.12$ & $0.235267-0.067425 i$ & $0.235267-0.067425 i$ & $0\%$\\
$0.6$ & $0.16$ & $0.241582-0.063618 i$ & $0.241582-0.063618 i$ & $0\%$\\
$0.6$ & $0.2$ & $0.249745-0.058524 i$ & $0.249745-0.058524 i$ & $0.\times 10^{\text{-4}}\%$\\
$0.6$ & $0.24$ & $0.259750-0.051958 i$ & $0.259747-0.051951 i$ & $0.0031\%$\\
$0.6$ & $0.28$ & $0.271513-0.043687 i$ & $0.271513-0.043687 i$ & $0\%$\\
$0.6$ & $0.32$ & $0.284792-0.033426 i$ & $0.284723-0.033547 i$ & $0.0487\%$\\
$0.6$ & $0.34$ & $0.291350-0.027954 i$ & $0.291439-0.028598 i$ & $0.222\%$\\
$1.$ & $0$ & $0.245302-0.079946 i$ & $0.245302-0.079946 i$ & $0\%$\\
$1.$ & $0.08$ & $0.248675-0.077951 i$ & $0.248675-0.077951 i$ & $0.\times 10^{\text{-4}}\%$\\
$1.$ & $0.12$ & $0.252908-0.075417 i$ & $0.252908-0.075417 i$ & $0.\times 10^{\text{-4}}\%$\\
$1.$ & $0.16$ & $0.258863-0.071787 i$ & $0.258863-0.071787 i$ & $0.\times 10^{\text{-4}}\%$\\
$1.$ & $0.2$ & $0.266562-0.066967 i$ & $0.266562-0.066967 i$ & $0\%$\\
$1.$ & $0.24$ & $0.276016-0.060819 i$ & $0.276016-0.060819 i$ & $0.\times 10^{\text{-4}}\%$\\
$1.$ & $0.28$ & $0.287197-0.053152 i$ & $0.287196-0.053153 i$ & $0.0005\%$\\
$1.$ & $0.32$ & $0.299993-0.043782 i$ & $0.299807-0.044315 i$ & $0.186\%$\\
$1.$ & $0.38$ & $0.330375-0.027939 i$ & $0.332519-0.026161 i$ & $0.840\%$\\
$1.4$ & $0$ & $0.255606-0.084114 i$ & $0.255606-0.084114 i$ & $0\%$\\
$1.4$ & $0.08$ & $0.258863-0.082173 i$ & $0.258863-0.082173 i$ & $0\%$\\
$1.4$ & $0.12$ & $0.262950-0.079710 i$ & $0.262950-0.079710 i$ & $0\%$\\
$1.4$ & $0.16$ & $0.268699-0.076191 i$ & $0.268699-0.076191 i$ & $0\%$\\
$1.4$ & $0.2$ & $0.276130-0.071535 i$ & $0.276130-0.071535 i$ & $0\%$\\
$1.4$ & $0.24$ & $0.285258-0.065623 i$ & $0.285258-0.065623 i$ & $0\%$\\
$1.4$ & $0.28$ & $0.296071-0.058291 i$ & $0.296071-0.058291 i$ & $0\%$\\
$1.4$ & $0.32$ & $0.308494-0.049336 i$ & $0.308492-0.049336 i$ & $0.0009\%$\\
$1.4$ & $0.36$ & $0.322318-0.038567 i$ & $0.322269-0.038603 i$ & $0.0186\%$\\
$1.4$ & $0.4$ & $0.354207-0.019828 i$ & $0.355643-0.017321 i$ & $0.814\%$\\
$1.8$ & $0$ & $0.262247-0.086697 i$ & $0.262247-0.086697 i$ & $0\%$\\
$1.8$ & $0.08$ & $0.265431-0.084792 i$ & $0.265431-0.084792 i$ & $0\%$\\
$1.8$ & $0.12$ & $0.269426-0.082378 i$ & $0.269426-0.082378 i$ & $0\%$\\
$1.8$ & $0.16$ & $0.275043-0.078933 i$ & $0.275043-0.078933 i$ & $0\%$\\
$1.8$ & $0.2$ & $0.282303-0.074383 i$ & $0.282303-0.074383 i$ & $0\%$\\
$1.8$ & $0.24$ & $0.291223-0.068621 i$ & $0.291223-0.068621 i$ & $0.\times 10^{\text{-4}}\%$\\
$1.8$ & $0.28$ & $0.301797-0.061496 i$ & $0.301797-0.061496 i$ & $0\%$\\
$1.8$ & $0.32$ & $0.313966-0.052823 i$ & $0.313968-0.052821 i$ & $0.0008\%$\\
$1.8$ & $0.36$ & $0.327608-0.042414 i$ & $0.327573-0.042400 i$ & $0.0114\%$\\
$1.8$ & $0.4$ & $0.342078-0.031883 i$ & $0.342868-0.032767 i$ & $0.345\%$\\
\hline\hline
\end{tabular}
\caption{\qnmCaption{1}{-0.3}}
\label{tab:ms-qnm-l1-p2m03}
\end{table}

\begin{table}[t]
\qnmTableSetup
\begin{tabular}{@{}c c c c c@{}}
\qnmHeader
$0.2$ & $0$ & $0.044845-0.033347 i$ & $0.043795-0.033492 i$ & $1.90\%$\\
$0.2$ & $0.04$ & $0.046025-0.030541 i$ & $0.046114-0.030521 i$ & $0.165\%$\\
$0.2$ & $0.08$ & $0.053689-0.015674 i$ & $0.052960-0.014319 i$ & $2.75\%$\\
$0.6$ & $0$ & $0.068328-0.060021 i$ & $0.068329-0.060037 i$ & $0.0172\%$\\
$0.6$ & $0.04$ & $0.068887-0.058009 i$ & $0.068890-0.058047 i$ & $0.0414\%$\\
$0.6$ & $0.08$ & $0.070972-0.052141 i$ & $0.070992-0.052139 i$ & $0.0224\%$\\
$0.6$ & $0.12$ & $0.071629-0.042768 i$ & $0.071299-0.044971 i$ & $2.67\%$\\
$0.6$ & $0.14$ & $0.088909-0.047499 i$ & $0.092631-0.043490 i$ & $5.43\%$\\
$1.$ & $0$ & $0.079110-0.072135 i$ & $0.078926-0.071958 i$ & $0.239\%$\\
$1.$ & $0.04$ & $0.079705-0.070283 i$ & $0.079410-0.070118 i$ & $0.318\%$\\
$1.$ & $0.08$ & $0.081168-0.064627 i$ & $0.081650-0.064954 i$ & $0.562\%$\\
$1.$ & $0.12$ & $0.082970-0.056684 i$ & $0.081885-0.058133 i$ & $1.80\%$\\
$1.$ & $0.14$ & $0.074342-0.048837 i$ & $0.082028-0.051791 i$ & $9.26\%$\\
$1.$ & $0.16$ & $0.079908-0.058963 i$ & $0.085624-0.062319 i$ & $6.67\%$\\
$1.4$ & $0$ & $0.085256-0.079077 i$ & $0.085172-0.078822 i$ & $0.231\%$\\
$1.4$ & $0.04$ & $0.085755-0.077426 i$ & $0.085623-0.077094 i$ & $0.309\%$\\
$1.4$ & $0.08$ & $0.087258-0.072423 i$ & $0.088327-0.072218 i$ & $0.960\%$\\
$1.4$ & $0.12$ & $0.088977-0.065355 i$ & $0.089014-0.064763 i$ & $0.537\%$\\
$1.4$ & $0.16$ & $0.086966-0.055294 i$ & $0.086913-0.055318 i$ & $0.0564\%$\\
$1.4$ & $0.18$ & $0.111852-0.062851 i$ & $0.117074-0.057653 i$ & $5.74\%$\\
$1.8$ & $0$ & $0.089395-0.083574 i$ & $0.089319-0.083302 i$ & $0.231\%$\\
$1.8$ & $0.04$ & $0.089861-0.081992 i$ & $0.089747-0.081648 i$ & $0.298\%$\\
$1.8$ & $0.08$ & $0.091340-0.077277 i$ & $0.093373-0.076184 i$ & $1.93\%$\\
$1.8$ & $0.12$ & $0.093177-0.070114 i$ & $0.093400-0.069618 i$ & $0.466\%$\\
$1.8$ & $0.16$ & $0.104096-0.051983 i$ & $0.092900-0.059674 i$ & $11.7\%$\\
$1.8$ & $0.2$ & $0.171765-0.000086 i$ & $0.171975-0.000063 i$ & $0.123\%$\\
\hline\hline
\end{tabular}
\caption{\qnmCaption{0}{-0.6}}
\label{tab:ms-qnm-l0-p2m06}
\end{table}

\begin{table}[t]
\qnmTableSetup
\begin{tabular}{@{}c c c c c@{}}
\qnmHeader
$0.2$ & $0$ & $0.121583-0.031960 i$ & $0.121583-0.031960 i$ & $0\%$\\
$0.2$ & $0.04$ & $0.123014-0.031236 i$ & $0.123014-0.031236 i$ & $0\%$\\
$0.2$ & $0.08$ & $0.127336-0.028971 i$ & $0.127336-0.028971 i$ & $0.0002\%$\\
$0.2$ & $0.12$ & $0.134583-0.024841 i$ & $0.134584-0.024842 i$ & $0.0006\%$\\
$0.2$ & $0.16$ & $0.144504-0.018212 i$ & $0.144512-0.018310 i$ & $0.0674\%$\\
$0.2$ & $0.18$ & $0.153754-0.013732 i$ & $0.154350-0.012796 i$ & $0.719\%$\\
$0.6$ & $0$ & $0.181264-0.056321 i$ & $0.181264-0.056321 i$ & $0\%$\\
$0.6$ & $0.04$ & $0.182369-0.055696 i$ & $0.182369-0.055696 i$ & $0\%$\\
$0.6$ & $0.08$ & $0.185693-0.053794 i$ & $0.185693-0.053794 i$ & $0\%$\\
$0.6$ & $0.12$ & $0.191266-0.050525 i$ & $0.191266-0.050525 i$ & $0\%$\\
$0.6$ & $0.16$ & $0.199120-0.045718 i$ & $0.199120-0.045718 i$ & $0.\times 10^{\text{-4}}\%$\\
$0.6$ & $0.2$ & $0.209245-0.039088 i$ & $0.209228-0.039077 i$ & $0.0096\%$\\
$0.6$ & $0.24$ & $0.221461-0.030255 i$ & $0.221447-0.030268 i$ & $0.0085\%$\\
$0.6$ & $0.28$ & $0.246866-0.013038 i$ & $0.247713-0.011332 i$ & $0.770\%$\\
$1.$ & $0$ & $0.209261-0.067381 i$ & $0.209261-0.067381 i$ & $0\%$\\
$1.$ & $0.04$ & $0.210241-0.066810 i$ & $0.210241-0.066810 i$ & $0\%$\\
$1.$ & $0.08$ & $0.213188-0.065079 i$ & $0.213188-0.065079 i$ & $0\%$\\
$1.$ & $0.12$ & $0.218123-0.062133 i$ & $0.218123-0.062133 i$ & $0.\times 10^{\text{-4}}\%$\\
$1.$ & $0.16$ & $0.225076-0.057869 i$ & $0.225076-0.057869 i$ & $0.\times 10^{\text{-4}}\%$\\
$1.$ & $0.2$ & $0.234066-0.052121 i$ & $0.234066-0.052121 i$ & $0\%$\\
$1.$ & $0.24$ & $0.245064-0.044640 i$ & $0.245064-0.044641 i$ & $0.0004\%$\\
$1.$ & $0.28$ & $0.257907-0.035118 i$ & $0.257902-0.035126 i$ & $0.0035\%$\\
$1.$ & $0.32$ & $0.273885-0.026065 i$ & $0.275491-0.026193 i$ & $0.586\%$\\
$1.4$ & $0$ & $0.225843-0.073737 i$ & $0.225843-0.073737 i$ & $0\%$\\
$1.4$ & $0.04$ & $0.226759-0.073196 i$ & $0.226759-0.073196 i$ & $0\%$\\
$1.4$ & $0.08$ & $0.229513-0.071561 i$ & $0.229513-0.071561 i$ & $0\%$\\
$1.4$ & $0.12$ & $0.234123-0.068789 i$ & $0.234123-0.068789 i$ & $0\%$\\
$1.4$ & $0.16$ & $0.240613-0.064801 i$ & $0.240613-0.064801 i$ & $0\%$\\
$1.4$ & $0.2$ & $0.249007-0.059471 i$ & $0.249008-0.059471 i$ & $0.0002\%$\\
$1.4$ & $0.24$ & $0.259306-0.052611 i$ & $0.259306-0.052611 i$ & $0\%$\\
$1.4$ & $0.28$ & $0.271433-0.043976 i$ & $0.271433-0.043972 i$ & $0.00140\%$\\
$1.4$ & $0.32$ & $0.285135-0.033319 i$ & $0.285088-0.033345 i$ & $0.0186\%$\\
$1.4$ & $0.34$ & $0.292078-0.027625 i$ & $0.292116-0.028143 i$ & $0.177\%$\\
$1.8$ & $0$ & $0.236877-0.077876 i$ & $0.236877-0.077876 i$ & $0\%$\\
$1.8$ & $0.04$ & $0.237755-0.077355 i$ & $0.237755-0.077355 i$ & $0\%$\\
$1.8$ & $0.08$ & $0.240392-0.075782 i$ & $0.240392-0.075782 i$ & $0\%$\\
$1.8$ & $0.12$ & $0.244804-0.073119 i$ & $0.244804-0.073119 i$ & $0\%$\\
$1.8$ & $0.16$ & $0.251013-0.069299 i$ & $0.251013-0.069299 i$ & $0\%$\\
$1.8$ & $0.2$ & $0.259043-0.064218 i$ & $0.259043-0.064218 i$ & $0\%$\\
$1.8$ & $0.24$ & $0.268903-0.057717 i$ & $0.268904-0.057717 i$ & $0.\times 10^{\text{-4}}\%$\\
$1.8$ & $0.28$ & $0.280551-0.049584 i$ & $0.280553-0.049584 i$ & $0.0008\%$\\
$1.8$ & $0.32$ & $0.293830-0.039588 i$ & $0.293866-0.039565 i$ & $0.0142\%$\\
$1.8$ & $0.36$ & $0.308062-0.028789 i$ & $0.308567-0.029560 i$ & $0.298\%$\\
\hline\hline
\end{tabular}
\caption{\qnmCaption{1}{-0.6}}
\label{tab:ms-qnm-l1-p2m06}
\end{table}

\begin{table}[t]
\qnmTableSetup
\begin{tabular}{@{}c c c c c@{}}
\qnmHeader
$0.2$ & $0$ & $0.200947-0.031789 i$ & $0.200947-0.031789 i$ & $0\%$\\
$0.2$ & $0.08$ & $0.204812-0.030647 i$ & $0.204812-0.030647 i$ & $0\%$\\
$0.2$ & $0.16$ & $0.216660-0.026969 i$ & $0.216660-0.026969 i$ & $0\%$\\
$0.2$ & $0.24$ & $0.237250-0.019607 i$ & $0.237250-0.019607 i$ & $0\%$\\
$0.2$ & $0.28$ & $0.250991-0.013459 i$ & $0.250868-0.013490 i$ & $0.0502\%$\\
$0.6$ & $0$ & $0.299132-0.055859 i$ & $0.299132-0.055859 i$ & $0\%$\\
$0.6$ & $0.08$ & $0.302210-0.054858 i$ & $0.302210-0.054858 i$ & $0\%$\\
$0.6$ & $0.16$ & $0.311522-0.051788 i$ & $0.311522-0.051788 i$ & $0\%$\\
$0.6$ & $0.24$ & $0.327314-0.046416 i$ & $0.327314-0.046416 i$ & $0\%$\\
$0.6$ & $0.32$ & $0.350020-0.038200 i$ & $0.350020-0.038200 i$ & $0\%$\\
$0.6$ & $0.4$ & $0.380112-0.025907 i$ & $0.380112-0.025907 i$ & $0.00005\%$\\
$0.6$ & $0.44$ & $0.398196-0.018461 i$ & $0.398753-0.018611 i$ & $0.145\%$\\
$1.$ & $0$ & $0.345386-0.066796 i$ & $0.345386-0.066796 i$ & $0\%$\\
$1.$ & $0.08$ & $0.348137-0.065877 i$ & $0.348137-0.065877 i$ & $0\%$\\
$1.$ & $0.16$ & $0.356437-0.063082 i$ & $0.356437-0.063082 i$ & $0\%$\\
$1.$ & $0.24$ & $0.370443-0.058275 i$ & $0.370443-0.058275 i$ & $0\%$\\
$1.$ & $0.32$ & $0.390429-0.051170 i$ & $0.390429-0.051170 i$ & $0\%$\\
$1.$ & $0.4$ & $0.416786-0.041185 i$ & $0.416786-0.041185 i$ & $0\%$\\
$1.$ & $0.44$ & $0.432465-0.034778 i$ & $0.432464-0.034778 i$ & $0.00025\%$\\
$1.$ & $0.48$ & $0.449788-0.027150 i$ & $0.449782-0.027149 i$ & $0.00150\%$\\
$1.$ & $0.52$ & $0.476252-0.013528 i$ & $0.476832-0.012055 i$ & $0.332\%$\\
$1.4$ & $0$ & $0.372793-0.073081 i$ & $0.372793-0.073081 i$ & $0\%$\\
$1.4$ & $0.08$ & $0.375372-0.072211 i$ & $0.375372-0.072211 i$ & $0\%$\\
$1.4$ & $0.16$ & $0.383145-0.069570 i$ & $0.383145-0.069570 i$ & $0\%$\\
$1.4$ & $0.24$ & $0.396236-0.065056 i$ & $0.396236-0.065056 i$ & $0\%$\\
$1.4$ & $0.32$ & $0.414858-0.058461 i$ & $0.414858-0.058461 i$ & $0\%$\\
$1.4$ & $0.4$ & $0.439328-0.049385 i$ & $0.439328-0.049385 i$ & $0\%$\\
$1.4$ & $0.44$ & $0.453867-0.043694 i$ & $0.453867-0.043694 i$ & $0\%$\\
$1.4$ & $0.48$ & $0.469980-0.037046 i$ & $0.469980-0.037046 i$ & $0.00004\%$\\
$1.4$ & $0.52$ & $0.487632-0.029249 i$ & $0.487627-0.029247 i$ & $0.00120\%$\\
$1.4$ & $0.56$ & $0.511321-0.020524 i$ & $0.512397-0.019428 i$ & $0.300\%$\\
$1.8$ & $0$ & $0.391032-0.077175 i$ & $0.391032-0.077175 i$ & $0\%$\\
$1.8$ & $0.08$ & $0.393505-0.076336 i$ & $0.393505-0.076336 i$ & $0\%$\\
$1.8$ & $0.16$ & $0.400955-0.073794 i$ & $0.400955-0.073794 i$ & $0\%$\\
$1.8$ & $0.24$ & $0.413487-0.069462 i$ & $0.413487-0.069462 i$ & $0\%$\\
$1.8$ & $0.32$ & $0.431286-0.063168 i$ & $0.431286-0.063168 i$ & $0\%$\\
$1.8$ & $0.4$ & $0.454624-0.054593 i$ & $0.454624-0.054593 i$ & $0\%$\\
$1.8$ & $0.44$ & $0.468473-0.049278 i$ & $0.468473-0.049278 i$ & $0\%$\\
$1.8$ & $0.48$ & $0.483823-0.043130 i$ & $0.483823-0.043130 i$ & $0\%$\\
$1.8$ & $0.52$ & $0.500679-0.035996 i$ & $0.500679-0.035996 i$ & $0.00006\%$\\
$1.8$ & $0.56$ & $0.518978-0.027689 i$ & $0.518967-0.027684 i$ & $0.00224\%$\\
$1.8$ & $0.58$ & $0.528514-0.023765 i$ & $0.528861-0.024143 i$ & $0.0971\%$\\
\hline\hline
\end{tabular}
\caption{\qnmCaption{2}{-0.6}}
\label{tab:ms-qnm-l2-p2m06}
\end{table}

\section{Greybody factors and absorption cross section}

For genuine scattering states with $\omega>\mu_s$, the same radial equation is instead solved with a unit-amplitude wave incident from spatial infinity,
\begin{equation}
\Psi \sim e^{-i\chi r_*} + R_{\ell} e^{+i\chi r_*}, \qquad r_*\to +\infty,
\end{equation}
\begin{equation}
\Psi \sim T_{\ell} e^{-i\omega r_*}, \qquad r_*\to -\infty,
\end{equation}
where $R_{\ell}$ and $T_{\ell}$ are the reflection and transmission amplitudes. The partial greybody factor is then
\begin{equation}
\Gamma_{\ell}(\omega) = |T_{\ell}|^2 = 1-|R_{\ell}|^2,
\end{equation}
and the total absorption cross section is obtained from the partial-wave sum
\begin{equation}
\sigma_{\rm abs}(\omega) = \sum_{\ell=0}^{\infty} \frac{\pi(2\ell+1)}{\chi^2}\,\Gamma_{\ell}(\omega).
\end{equation}
In this language, the greybody factor measures the fraction of incident flux transmitted through the curvature barrier, while the absorption cross section packages the partial-wave transmission probabilities into a single effective area. 

A useful by-product of the quasinormal-mode analysis is that it already contains much of the information needed to sketch the greybody factors, because in the WKB picture both the resonant frequencies and the transmission probabilities are controlled by the same local characteristics of the effective barrier near its maximum~\cite{Konoplya:2019hlu,Konoplya:2026wkb,Han:2026fpn}. In practice, the fundamental mode for a given multipole provides a compact summary of the corresponding transmission curve: ${\rm Re}(\omega_0)$ determines the frequency scale at which the partial greybody factor changes from strong reflection to substantial transmission, while $|{\rm Im}(\omega_0)|$ controls how sharp or smooth this transition is. Using the leading WKB relations between the barrier parameters and the fundamental frequency, one may write the transmission coefficient in the following form \cite{Konoplya:2024lir,Konoplya:2024vuj}
\begin{equation}
\label{eq:ms-gbf-qnm-correspondence}
\Gamma_{\ell}(\omega) \approx \left[1 + \exp\!\left(2\pi\frac{{\rm Re}(\omega_0)^2-\omega^2}{4\,{\rm Re}(\omega_0)\,|{\rm Im}(\omega_0)|}\right)\right]^{-1} + \mathcal{O}(\ell^{-1}),
\end{equation}
which makes the correspondence especially transparent in the eikonal regime. The correspondence has been applied and tested across a wide range of configurations, demonstrating good accuracy already at moderate values of $\ell$ \cite{Malik:2024cgb,Lutfuoglu:2025mqa,Skvortsova:2024msa,Bolokhov:2025lnt,Konoplya:2010vz,Lutfuoglu:2026uzy,Malik:2025dxn,Dubinsky:2024vbn,Lutfuoglu:2025eik,Bolokhov:2024otn,Malik:2024wvs,Lutfuoglu:2025kqp,Bolokhov:2026eqf,Lutfuoglu:2025ldc}. It is known to break down in situations where the WKB method itself ceases to be applicable for determining quasinormal modes, such as in the presence of double-well effective potentials or in certain higher-curvature theories that give rise to eikonal instabilities \cite{Konoplya:2025hgp,Bolokhov:2023dxq,Konoplya:2017ymp}.

In the present analysis the greybody factors, and consequently the absorption cross sections discussed below, are reconstructed solely from the quasinormal-mode data through the correspondence formula~\eqref{eq:ms-gbf-qnm-correspondence}, rather than by a direct numerical integration of the scattering problem. Therefore, the following discussion should be understood as an interpretation of the semiclassical trends encoded in the QNM spectrum and in the barrier characteristics to which Eq.~\eqref{eq:ms-gbf-qnm-correspondence} is sensitive.

\begin{figure*}[t]
    \centering
    \begin{minipage}{0.315\textwidth}
        \centering
        \includegraphics[width=\linewidth]{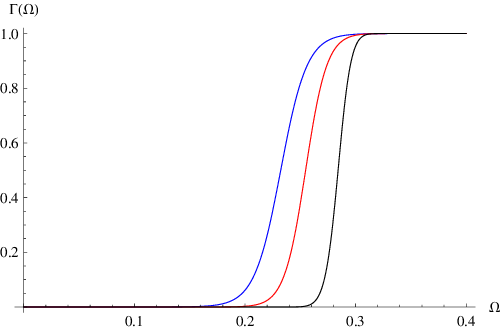}\\[-0.5ex]
        {\small (a) $\ell=2$}
    \end{minipage}\hfill
    \begin{minipage}{0.315\textwidth}
        \centering
        \includegraphics[width=\linewidth]{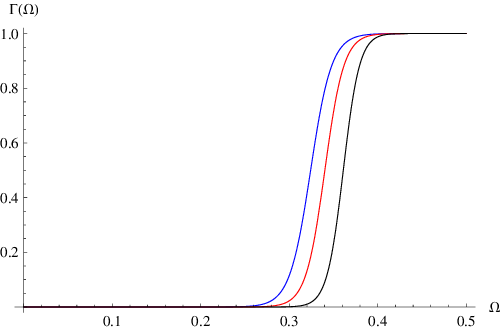}\\[-0.5ex]
        {\small (b) $\ell=3$}
    \end{minipage}\hfill
    \begin{minipage}{0.315\textwidth}
        \centering
        \includegraphics[width=\linewidth]{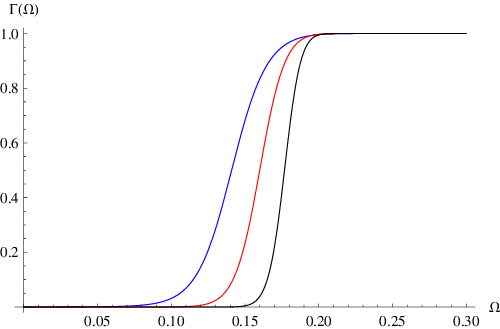}\\[-0.5ex]
        {\small (c) $\ell=1$}
    \end{minipage}
    \caption{Representative partial greybody factors $\Gamma_{\ell}(\omega)$ for the parity-symmetric beyond-Horndeski black hole with $M=1$, $p_1=0.2$, and $p_2=-0.5$. Panel (a) shows the $\ell=2$ curves for $\mu_s=0$ (blue), $\mu_s=0.2$ (red), and $\mu_s=0.3$ (black). Panel (b) shows the corresponding $\ell=3$ curves for the same masses, while panel (c) shows the $\ell=1$ case for $\mu_s=0$ (blue), $\mu_s=0.15$ (red), and $\mu_s=0.2$ (black). In all three panels, increasing the field mass shifts the transmission curve toward higher frequencies.}
    \label{fig:ms-gbf-representative}
\end{figure*}

Figure~\ref{fig:ms-gbf-representative} shows that, at fixed multipole number, increasing the scalar-field mass lowers the partial greybody factors over the transition region and shifts the onset of efficient transmission to higher frequencies. In each panel of Fig.~\ref{fig:ms-gbf-representative}, the smallest mass corresponds to the leftmost step and therefore to the largest transmission at a given frequency, while the larger masses move the step to the right and suppress the low- and intermediate-frequency part of the spectrum. In this sense, increasing $\mu_s$ filters out progressively softer radiation and transfers the main transmitted flux to higher frequencies. This is precisely the trend expected from Eq.~\eqref{eq:ms-gbf-qnm-correspondence}: when the characteristic QNM frequency scale grows, the logistic transition in $\Gamma_{\ell}(\omega)$ is displaced toward larger $\omega$.

The comparison among panels~(a)--(c) of Fig.~\ref{fig:ms-gbf-representative} also shows that higher multipoles have smaller greybody factors and require larger frequencies before the transmission approaches unity. Physically, this is consistent with the effective-potential picture, because increasing $\ell$ raises the centrifugal contribution and hence makes the barrier taller. A taller barrier is less transparent, so the corresponding partial wave is more strongly reflected and its greybody factor remains smaller until the incident frequency becomes sufficiently large. Within the QNM-based reconstruction, the same statement is encoded by the larger characteristic frequencies associated with higher-$\ell$ modes, which again push the transition region of Fig.~\ref{fig:ms-gbf-representative} to the right. Thus the mass dependence and the multipole dependence of the GBFs are both consistent with the barrier interpretation and with the QNM correspondence formula~\eqref{eq:ms-gbf-qnm-correspondence}.

\begin{figure*}[t]
    \centering
    \begin{minipage}{0.475\textwidth}
        \centering
        \includegraphics[width=\linewidth]{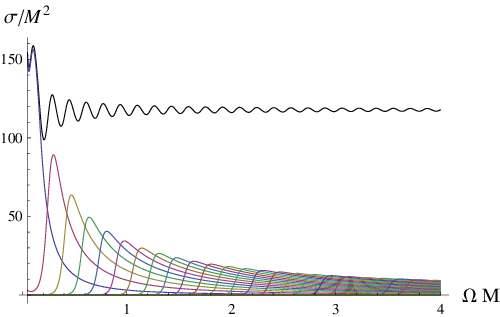}\\[-0.5ex]
        {\small (a) $p_1=2$}
    \end{minipage}\hfill
    \begin{minipage}{0.475\textwidth}
        \centering
        \includegraphics[width=\linewidth]{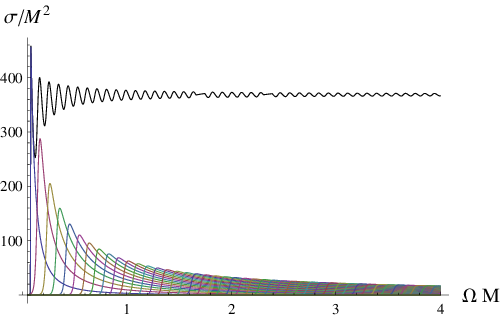}\\[-0.5ex]
        {\small (b) $p_1=0.2$}
    \end{minipage}
    \caption{Representative total and partial absorption cross sections for the massive scalar field in the parity-symmetric beyond-Horndeski black-hole background with $M=1$, $p_2=-0.5$, and $\mu_s=0.1$. Panel (a) corresponds to the near-Schwarzschild case $p_1=2$, while panel (b) corresponds to the more strongly deformed case $p_1=0.2$. In each panel the black curve shows the total absorption cross section $\sigma_{\rm abs}/M^2$, whereas the colored curves show the individual partial-wave contributions. The comparison indicates that moving away from the near-Schwarzschild regime enhances the overall absorption and increases the partial contributions over most of the displayed frequency range.}
    \label{fig:ms-acs-representative}
\end{figure*}

A complementary view is provided by the absorption cross section itself, shown in Fig.~\ref{fig:ms-acs-representative}. Since $\sigma_{\rm abs}$ is built from the partial transmission probabilities, the trends seen in Fig.~\ref{fig:ms-gbf-representative} must be reflected in the total and partial absorption profiles as well. In both panels of Fig.~\ref{fig:ms-acs-representative} the colored curves represent the separate partial-wave contributions, while the black curve is the total cross section obtained from their weighted sum. Because the GBFs entering this construction are inferred exclusively from Eq.~\eqref{eq:ms-gbf-qnm-correspondence}, Fig.~\ref{fig:ms-acs-representative} should likewise be interpreted as a QNM-based reconstruction of the absorption spectrum.

Figure~\ref{fig:ms-acs-representative} compares two cases with the same $M=1$, $p_2=-0.5$, and $\mu_s=0.1$, but different values of $p_1$. Since the Schwarzschild limit is approached by increasing $p_1$ at fixed $p_2$, panel~(a) with $p_1=2$ is the near-Schwarzschild case, whereas panel~(b) with $p_1=0.2$ is more strongly deformed. The comparison shows that the more strongly deformed case has systematically larger partial contributions and a noticeably higher total absorption cross section over the plotted frequency range. Therefore, within the present QNM-correspondence approach, moving away from the near-Schwarzschild regime enhances the absorption efficiency, while approaching Schwarzschild lowers it. The same conclusion is compatible with the behavior of the partial GBFs in Fig.~\ref{fig:ms-gbf-representative}: once the transmission curves are shifted and amplified in the more strongly deformed case, the summed absorption profile is enhanced accordingly.

\section{Conclusions}

In this work we formulated the perturbation and scattering problems for a minimally coupled massive scalar field on the asymptotically flat parity-symmetric beyond-Horndeski black-hole background. Starting from the exact metric function, we derived the massive Klein--Gordon equation, reduced it to a Schr\"odinger-type radial equation, and identified the effective potential that controls quasinormal oscillations, transmission through the curvature barrier, and the absorption cross section. A central qualitative result is that the scalar mass does not merely shift the spectrum: it also changes the shape of the potential at large radii by lifting it to the asymptotic plateau $V_{\ell}\to\mu_s^2$, and for representative parameter choices it can even remove the usual local barrier maximum. This feature explains why the massive problem becomes qualitatively different from the massless one near the threshold regime.

To study the spectrum we combined Pad\'e-improved high-order WKB calculations with characteristic time-domain evolution and Prony extraction. The comparison between WKB14 and WKB16 provides a useful internal accuracy diagnostic, and it shows a clear hierarchy across multipoles. The $\ell=1$ and $\ell=2$ modes remain very stable over most of the parameter range, typically with inter-order differences well below one percent, whereas the $\ell=0$ sector is significantly more delicate and becomes least reliable precisely when the barrier is shallow. The time-domain signal confirms this picture: for the representative mode $\ell=1$, $p_1=0.2$, $p_2=-0.3$, and $\mu_s=0.16$, the Prony frequency agrees with the WKB16--Pad\'e result to $0.024\%$ in the real part and $0.263\%$ in the imaginary part. The evolution also shows the expected late-time crossover from exponentially damped ringing to an oscillatory massive tail, whose envelope is approximately $t^{-1}$ in the displayed example.

The main spectral trend revealed by the tables is the pronounced decrease of the damping rate as $\mu_s$ increases. For several representative sequences the imaginary part of the frequency is driven toward zero in a systematic way, and the extrapolated values $\mu_{\rm crit}\approx0.229$, $0.303$, $0.549$, and $0.685$ suggest the onset of very long-lived quasi-resonant behavior. At the same time, these extrapolations should be interpreted cautiously: the same near-threshold regime is also the one in which the effective barrier becomes less standard and the WKB assumptions are under the greatest stress. Another robust trend is the close correlation between ${\rm Re}(\omega)$ and the barrier height, especially for $\ell\geq1$, which confirms that the oscillation frequency continues to track the local geometry of the effective potential even after the mass term substantially modifies the outer region.

On the scattering side, we used the QNM/barrier correspondence to reconstruct the qualitative behavior of the greybody factors and absorption cross section from the same effective potential. Within this semiclassical description, increasing $\mu_s$ shifts the onset of transmission to higher frequencies and suppresses the low- and intermediate-frequency greybody factors, while increasing $\ell$ makes the barrier less transparent and delays the approach to unit transmission. The absorption spectra inherit the same behavior. In particular, for fixed $M=1$, $p_2=-0.5$, and $\mu_s=0.1$, the more strongly deformed case $p_1=0.2$ exhibits larger partial contributions and a higher total absorption cross section than the near-Schwarzschild case $p_1=2$. These trends indicate that the massive scalar sector is sensitive not only to the field mass itself but also to how far the geometry lies from the Schwarzschild limit.

Overall, the present analysis shows that a massive scalar field provides a sharper probe of the parity-symmetric beyond-Horndeski black hole than its massless counterpart. The field mass competes with the geometric deformation parameters in the ringing regime, in the late-time tail, and in the transmission properties of the curvature barrier. An important practical lesson is that the frequency-domain and time-domain pictures need not remain equally transparent in the large-mass regime: even when the QNM spectrum points to very weak damping, the corresponding long-lived mode may be hidden in the time-domain signal because the massive tail takes over too early. In the Koyama--Tomimatsu picture, this crossover occurs on the scale $\mu_s t\gg \mu_s M$ for sufficiently large $\mu_s M$, so for the present normalization $M=1$ the onset of tail dominance can remain relatively early even while the QNM damping rate is already strongly suppressed. The most important next steps are a direct numerical integration of the scattering problem, a dedicated study of quasibound and near-threshold states, and a broader parameter survey extending the time-domain analysis deeper into the weak-damping regime. These developments would clarify the precise status of the quasi-resonant behavior suggested here and would place the greybody and absorption results on a fully independent numerical footing.

\vspace{5mm}
\begin{acknowledgments}
The author would like to acknowledge R. A. Konoplya for helpful discussions.
\end{acknowledgments}

\bibliographystyle{apsrev4-1}
\bibliography{refs}

\end{document}